# Non-solitary wave response generated by "soft" striker impact on strongly nonlinear weakly dissipative granular chains

Po-Hsun Chiu, Vitali F. Nesterenko*

Department of Mechanical and Aerospace Engineering, and Materials Science and Engineering Program, University of California at San Diego,

La Jolla, CA 92093-0411, USA

*Corresponding Author: Vitali F. Nesterenko (vnesterenko@ucsd.edu)

## Abstract

It is well known that striker impact on strongly nonlinear, non-dissipative or weakly dissipative granular chains assembled from spheres (or cylinders with flat ends and spheres) having similar masses results in strongly nonlinear Nesterenko solitary waves over a very short distance from the impacted end. This paper presents the first experimental results on the nature of the wave response excited by striker impact on strongly nonlinear weakly dissipative discrete chains assembled from cylinders with flat ends and spheres having similar masses depending on the ramp time of the incoming pulse. Variation of the pulse ramp time was accomplished using "hard" and "soft" striker impact. In the case of "hard" impact, the incoming pulse with relatively short ramp time was split into Nesterenko solitary waves after traveling about 20 particles as in previous publications. The "soft" impact by the same striker with the same velocity was achieved by placing a small mass PTFE spherical particle inside a PTFE O-ring at the top of the first cylinder resulting in significantly longer ramp time of the incoming pulse. The incoming pulse generated by "soft" impact did not split into a train of solitary waves despite the initial steepening of the leading front due to strong nonlinearity. Instead, it traveled as a shock-like stress wave a long distance of 131 particles. Thus, the generation of a train of solitary waves or shock-like stress wave in a weakly dissipative, strongly nonlinear discrete chain depends on the duration of the incoming pulse ramp. Ramping the incoming pulse resulted in a dramatic decrease of shock-like stress wave amplitude in comparison with the amplitude of the leading solitary wave in the case of "hard" impact.

## Introduction

The Los Alamos Scientific Laboratory report by Fermi, Pasta, and Ulam [1] was "dealing with the behavior of certain nonlinear physical systems where the non-linearity is introduced as a perturbation to a primarily linear problem" with "the primary aim of establishing, experimentally, the rate of approach to the equipartition of energy among the various degrees of freedom of the system". They reported that "the results show very little, if any, tendency toward equipartition of energy among the degrees of freedom". This result, at that time looked paradoxical, contradicting the generally accepted paradigm probably formulated first by D. Bernoulli in 1741: "for the elongation will not be proportional to the extending force ... and everything must be irregular" [2]. Attempts to solve this paradox started research into nonlinear wave dynamics in discrete systems. Zabusky [3] derived a long wave approximation (Korteweg-de Vries (KdV) equation) for a weakly nonlinear lattice, which has the solitary wave as the stationary solution. It connected earlier research on solitary waves in hydrodynamics, traced to an amazing experimental discovery by Russell [4], [5] of waves of translation (solitary waves) and to theoretical works by Boussinesq [6], Korteweg and de Vries [7]. Zabusky and Kruskal [8] coined the term soliton applied to the stationary solution of the KdV equation, which was subsequently adopted in many areas of physics [9], probably except for the mechanics of materials area.

The landscape of nonlinear wave dynamics has shifted to the mechanics of materials area after the publication of two papers—the theoretical paper by Nesterenko [10] and the experimental paper by Lazaridi and Nesterenko [11]. These papers addressed very different practically important physical problems than in papers [1], [3], [8]: damping high-amplitude pulses created by the impact or contact explosion. Investigation of wave propagation in a one-dimensional granular chain composed of spherical particles [10], [11], [12] was the first attempt to understand the damping properties of granular beds to impact or contact explosion. But unlike in examples considered in papers [1], [3], [8], interaction between spherical grains (or in physically similar granular chains assembled from spherical grains and cylinders with flat ends investigated in this article) is described by the strongly nonlinear Hertz law [13], [14]. This law for contact interaction between masses has no linear component if the chain is not precompressed. The initial static precompression of the chain introduces the linear component of the contact force, which has a negligible effect if the amplitude of the dynamic perturbation is much larger than the force at static precompression. Thus, in these cases, the nonlinearity cannot be introduced "as a perturbation to a primarily linear problem". That is why it was defined as a strongly, or essentially nonlinear problem [10], [12], [15], [16], [17].

Strongly nonlinear interactions between elastic spherical particles in a one-dimensional granular chain required the development of qualitatively new wave dynamics different from the behavior of chains with elastically linear or weakly nonlinear interactions between elements [12], [15-20]. Granular systems are probably the only structures that allow easy tuning (by applying precompression static force) of their behavior from strongly nonlinear to weakly nonlinear and to linear. These chains, when not precompressed by a static force, can be characterized as "sonic vacuum" (term coined by Nesterenko in [16], [17]), where the sound speed in its classical sense (the speed of long wave disturbances with negligible amplitude)) approaches zero. Unlike chains with weakly nonlinear interactions, a chain in “sonic vacuum” state supports strongly nonlinear solitary waves—Nesterenko solitary waves, a term coined by Coste, Falcon and Fauve [21] who independently confirmed the experimental results first presented in [11]. The strongly nonlinear solitary waves in “sonic vacuum” were first discovered as an exact stationary solution of a strongly nonlinear wave equation in the long-wave approximation first introduced in [10], which is more general than the weakly nonlinear KdV equation. The KdV wave equation is a limiting case of the strongly nonlinear wave equation when strains in the wave are much smaller than the initial strain due to precompression [18], [19]. It is well known that strongly nonlinear solitary waves are qualitatively different from weakly nonlinear solitary waves in a discrete chain and from corresponding KdV solitary waves [19]. For example, the spatial size of Nesterenko solitary waves in “sonic vacuum” is independent of wave amplitude, unlike the case of KdV solitary waves or weakly nonlinear waves in a discrete system. These waves behave like quasiparticles with an effective mass [22], and their more accurate description, compared with the exact solution of the long-wave approximation [10], can be found in papers by Sen and Manciu [23], [24] and Chatterjee [25]. Weakly nonlinear systems require much longer distances for the formation of solitary waves; thus, experimental observation of KdV solitary waves is possible only at very low dissipation, preventing wave attenuation. Recent insights into why the KdV solution for granular chains is not observed were presented in [26].

Discrete strongly nonlinear mechanical systems allow not only experimental investigation of wave dynamics in all regimes (from “sonic vacuum”/strongly nonlinear to linear cases) but also guide the design of new metamaterials for impact mitigation and wave processing, arranging various elements in different geometries. Many aspects of traditional wave dynamics should be reexamined in the case of "sonic vacuum." For example, wave interaction with the interfaces of two "sonic vacua" results in an unusual phenomenon exhibiting "acoustic diode" behavior [27]. Classical Snell's law cannot be applied due to the absence of sound speed; instead, the reflection and transmission laws of solitary waves are governed by their speeds [28].

It should be emphasized that a granular chain with Hertzian interaction law was the first theoretically and experimentally investigated example of strongly nonlinear discrete media. Other solitary and shock-like stress waves in discrete systems were also analyzed for the following interaction law between masses:

(a) power law interaction between particles (normal behavior, exponent larger than 1) [12], [15], [16], [17];

(b) abnormal power law (exponent less than 1) [20], [29];

(c) double power law, experimentally realizable using O-rings between cylinders with flat ends. It is even more strongly nonlinear than the classical Hertzian law between particles with two terms: Hertzian term with exponent 3/2 and additional term with exponent 6 [30];

(d) A complex interaction law represented by piecewise smooth arcs, arrangements of strongly nonlinear O-rings with inserted smaller diameter O-rings, or a nonlinear elastic layer between masses. This arrangement allows a broad combination of strong nonlinearity by placing smaller diameter O-rings inside larger diameter O-rings, or placing nonlinear elastically softening layers inside O-rings, combining "normal" and "abnormal" strong nonlinearity [31].

(e) general interaction law [18], [19], [32], [33].

Strongly nonlinear discrete materials can also be assembled based on complex metastructures [34]. Other examples are transverse motion of an elastic string with zero tension [18], [35] or using tensegrity cells [34].

Periodic nonlinear waves in a precompressed discrete chain with Hertzian interaction were analyzed numerically, and results agreed with the approach based on long wave approximation in [37].

The band gap in strongly nonlinear two-mass chains, consisting of particles with different masses arranged alternately, can be effectively tuned by static precompression [38]. They exhibit novel behavior, supporting solitary waves in the "sonic vacuum" only at a specific mass ratio of the particles [39].

The fact that dissipation in steel granular chains (e.g., friction between grains and between grains and supporting structure) was weak allowed the first experimental observation of Nesterenko solitary waves [11] and their observation by other researchers with particles made from different materials [21], [40-44]. It should be mentioned that dissipation was not considered in the theoretical approach and in the numerical calculations, presenting the discovery of a strongly nonlinear Nesterenko solitary wave in [10]. Validation of the existence of strongly nonlinear solitary waves in experiments was crucial, and their existence was only possible at weak dissipation. The critical value of dissipation rate preventing the generation of solitary waves in "sonic vacuum" is not known, but it is expected to be like a critical value of effective viscous dissipation related to a transition from oscillatory to monotonous shock waves found in [45], [46]. The high level of dissipation prevents the generation of solitary waves in some granular chains despite the discrete nature and strong nonlinearity. For example, granular chains assembled using O-rings between cylinders with flat ends have nonlinearity stronger than the classical Hertzian law with two terms (Hertzian term with exponent 3/2 and additional term with larger exponent 6). But they did not support strongly nonlinear solitary waves in experiments [47].

One of the remarkable features of discrete "sonic vacuum" systems is their ability to transform dynamic input caused by striker impact (or by contact explosion) into a train of Nesterenko solitary waves over relatively short distances from the impacted end [10], [40-44]. The number of solitary waves depends on the incoming pulse duration, determined by the mass of the striker or the duration of the incoming pulse. This unique feature was explored in many publications using the impact generation of solitary waves in discrete systems.

A practically important aspect of striker impact on a granular chain is that the striker does not recoil after impact. This is a unique feature which is very important when granular beds (Fig. 4 in [19]) are used to support explosive working (e.g., welding, shock compression). It was explained by the very low acoustic impedance of granular beds (which is zero if no precompression is applied – "sonic vacuum" case) compared with that of the striker (or metal plates on top of the granular bed under contact explosion) [19]), despite the acoustic impedance of iron grains in granular beds being like that of metal plates.

This article addresses the engineering problem of enhancing the mitigation of striker impact. We experimentally demonstrated that an identical striker with the same velocity generates two significantly different ramp times of the incoming pulse due to the placement of a "soft" PTFE sphere at the top of the chain. It was the first experimental observation that ramping of the leading part of the incoming pulse, created by the same striker with the same velocity, from 30 microseconds ("hard" impact) to 250 microseconds ("soft" impact), dramatically changes the nature of the propagated wave in the strongly nonlinear, weakly dissipative discrete system. In the case of "soft" impact, the incoming pulse was transformed into the shock-like stress wave propagating over relatively long distances, while “hard” impact resulted in the train of Nesterenko solitary waves. The amplitude of the shock-like stress wave generated by “soft” impact was dramatically smaller than the amplitudes of solitary waves in the case of “hard” impact. This experimental observation reveals a crucial role of the incoming pulse ramping time in the wave response of the granular chain.

**Experiments**

Experiments were conducted with chains composed of SS304 stainless steel cylinders (McMaster-Carr) and 440C stainless steel balls having a similar mass, with the arrangement as in [14]. Forces between cylinders and spheres follow Hertz law [13], [14] as in the chain assembled from spheres, the difference is only in the constants in Hertz law. Thus, these two granular systems are similar with respect to wave propagation. Granular chain composed from cylinders with flat ends and spheres was selected due to the following reasons: (a) it supports strongly nonlinear solitary waves similar to the chain composed from spheres [14]; (b) this system can be analyzed using experimentally validated dissipative numerical model [14]; (c) sensors embedded into cylinders are well aligned with the axis of granular chain and do not lose the alignment after each impact. This is especially important for reproducibility of results when multiple sensors are used in experiments, as is the case in our article; (d) it allows experimental study of granular systems with arbitrary mass ratio without changing contact interaction [14].

It should be emphasized that the wave dynamics of granular chains are governed by a contact interaction law depending on the radius of contacting surfaces, the elastic properties of particles, particle masses, and the distance between their centers. Overall shape of particles does not play any role in wave dynamics when the characteristic time of wave propagation inside particles is much smaller than the characteristic time of propagating wave (e.g., duration of solitary wave) [10], [12], [18].

The chain was placed vertically inside the channel made by four aluminum alloy rods arranged in a square pattern. They provided alignment of balls and cylinders with minimal contact with the supporting rods, minimizing dissipation due to friction. The steel cylinders and spheres have the same diameter, 8 mm and almost identical masses ($m$=2.05 g for cylinders and $m$=2.08 g for spheres). The height of the steel cylinder was $h$ = 5.3 mm.

The piezoelectric sensors were cut from the piezoelectric ceramic plate, supplied by Piezo Systems, Inc., wired and embedded between two half-cylinders using epoxy, ensuring a similar mass and height to the other cylinders in the chain, like in papers [14], [30].

The piezoelectric sensors were calibrated using the impact of a 440C steel sphere (mass 2.08 g) based on linear momentum conservation, which gives the value of the force integral over time from the start of the impact to the maximum force detected by the sensor. The results of 3 calibration experiments with Sensor 1 with the same striker impact are presented in Fig. 1.

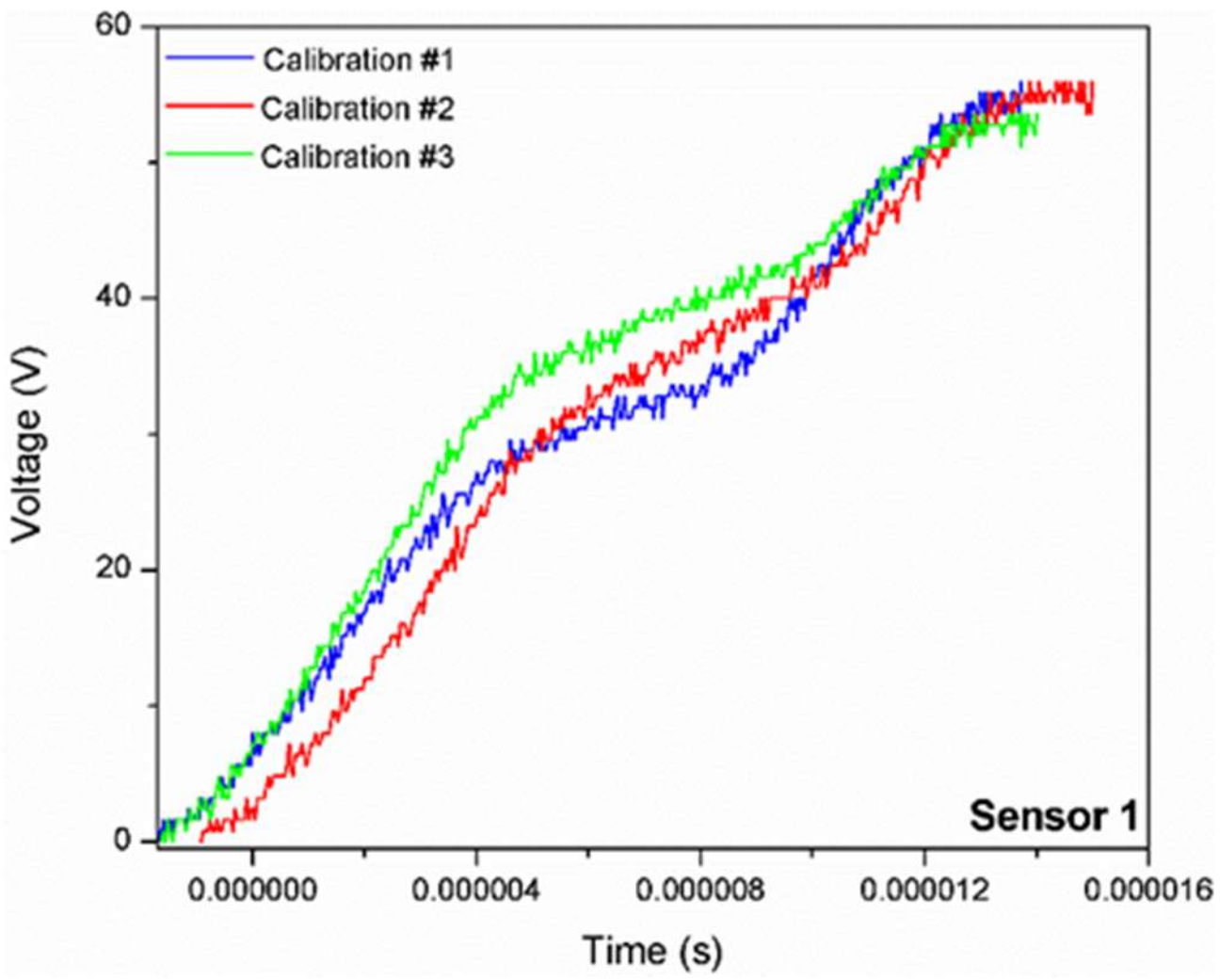

Figure 1. Calibration records of Sensor 1 under impact by one bead with mass the same as in the chain and an impact velocity of 0.83 m/s. The calibration factor $k$ for Sensor 1 is 4.84 (N/V).

Assuming a linear relation between force and voltage generated by sensors, calibration factors k = 4.84, 13.25, 21.02, and 7.08 (N/V) were obtained, respectively, for Sensors 1, 2, 3, and 4.

In experiments, 150 particles (cylinders and spheres) were arranged into a one-dimensional chain. Four sensors were placed at different positions along the chain to monitor the shape of the incoming pulses and their evolution during the propagation. Sensor 1 is located at the first particle (cylinder) from the top, Sensors 2, 3 and 4 are located at the 21st, 121st and 131st particles from the top, respectively. Sensor 4 is placed 20 particles from the bottom, Fig. 2.

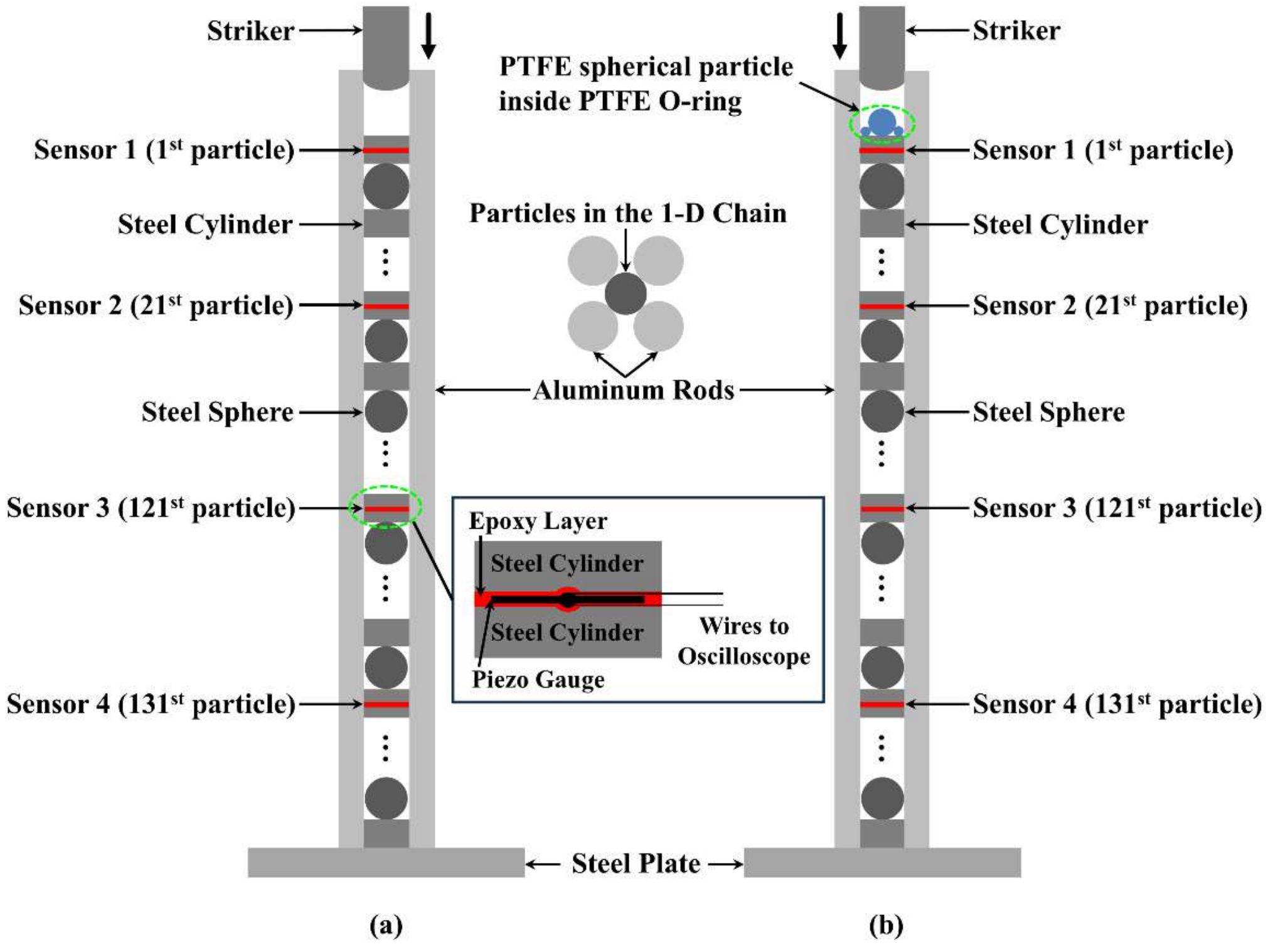

Figure 2. Two experimental setups resulting in different ramp times of incoming pulses: “hard” impact (Fig. 1 (a)) and “soft” impact (Fig. 1 (b)). The embedded piezo gauge inside the steel cylinder was shown closer to the bottom in the central part of the figure.

The waves in the systems were generated by the impact of a steel striker, diameter = 7.9 mm with a rounded tip. The length of the striker was 26.8 mm; its mass was 10.3 g, about five times the mass of a steel sphere or a cylinder. The signals from the sensors in “hard” and “soft” impact experiments were recorded using a Tektronix TDS 2014 oscilloscope. The trigger source was the signal coming from the channel connected to Sensor 1 (the first particle), configured with a Rising Edge trigger. The trigger threshold was set to a voltage corresponding to the initial contact force, allowing all four channels to be recorded simultaneously once triggered by Sensor 1. This setup provided consistent and repeatable records generated by all 4 sensors in each experiment.

Responses to two types of impact loading were investigated with similar discrete systems. In the first, the striker impacted the top steel cylinder directly, called a “hard” impact (Fig. 2(a)). The second arrangement (Fig. 2(b)), called a “soft” impact, was identical to the first one except that the same striker with the same velocity impacted the top PTFE sphere, creating a ramped incoming pulse. PTFE sphere (diameter: 4.8 mm) was put inside a PTFE O-ring (O-Rings West, O.D.: 7.9 mm, I.D.: 4.8 mm) to ensure the central position of the PTFE sphere on top of the cylinder with Sensor 1. The PTFE sphere mass (0.1226 g) was about 84 times smaller than the striker mass. The top of the PTFE sphere was initially 1.7 mm above the top of the O-ring; thus, only contact deformation between the PTFE sphere and striker controls ramp time, amplitude, and duration of the incoming stress wave. We would like to mention that the PTFE O-ring was used only to position the PTFE ball at the center of the top cylinder and did not play a significant role in striker interaction with the PTFE ball. PTFE O-ring helps to ensure one-dimensional geometry of the investigated configuration, keeping the PTFE sphere in the central position and restricting sphere’s side movements during impact. We suggest that in future numerical modeling PTFE O-ring can be neglected. Dynamic elastic properties of contact deformation of PTFE spheres are significantly different than their static properties as discussed in [47].

The steel striker with about five times the mass of a particle was dropped from 50 mm above the first particle of the 1-D chain, resulting in an impact velocity close to 1 m/s. The shape and amplitudes of waves for “hard” and “soft” impact are presented in Fig. 3. They are dramatically different: “hard” impact generated a train of solitary waves, while “soft” impact resulted in a shock-like stress wave with smaller amplitude.

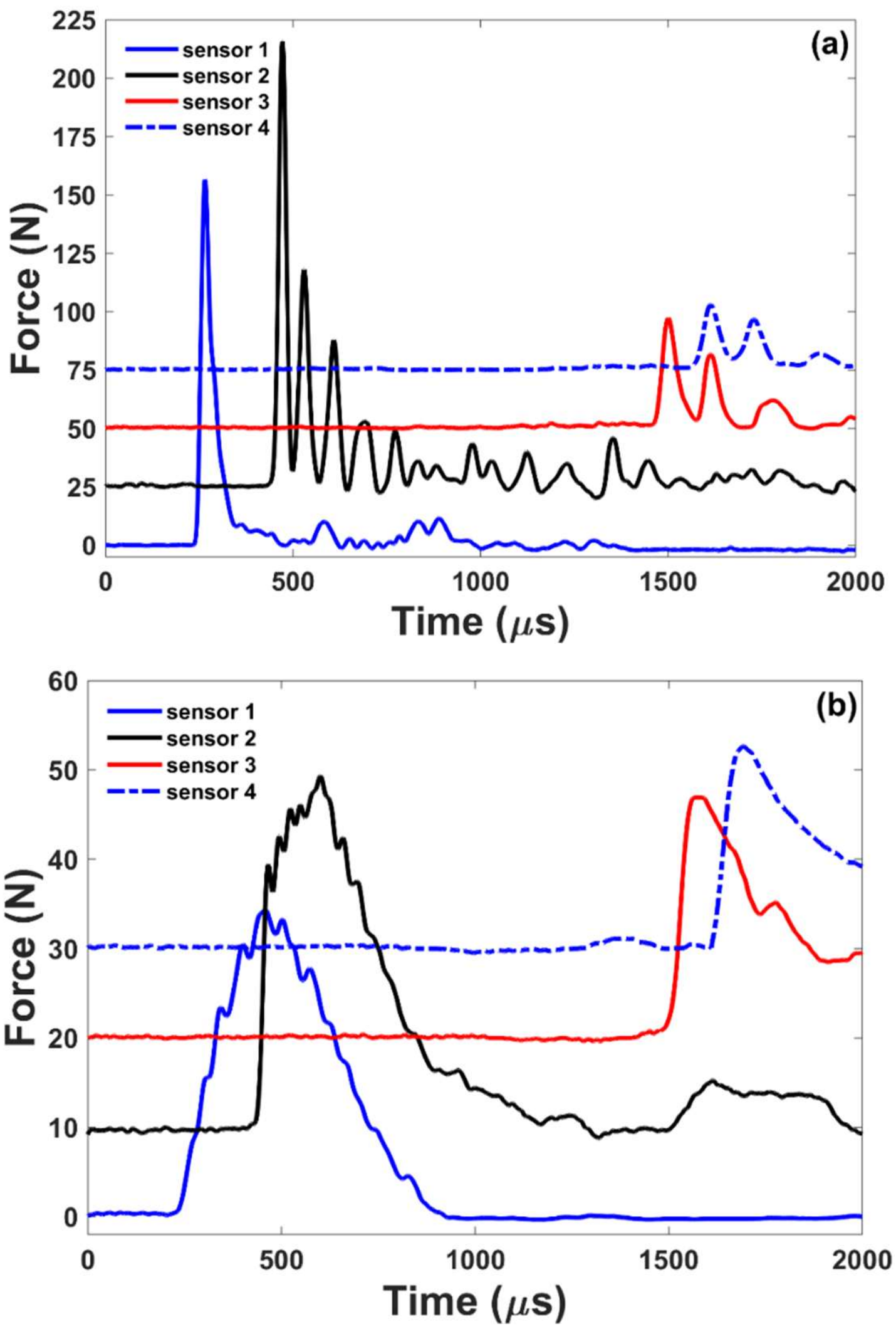

Figure 3. Waves excited by a striker with a velocity of 1 m/s: a solitary wave train excited by "hard" (a) and shock-like stress wave excited by "soft" (b) impacts. The curves corresponding to Sensors 1-4 are offset by 25 N (a) and 10 N (b) for visual clarity.

Dissipation due to friction between cylinders and spheres and between them and supporting rods can depend on wave amplitudes. Thus, we investigated the nature of the wave response excited by "hard" impact when the amplitude of the incoming pulse was comparable to the shock-like stress wave amplitude excited by "soft" impact (Fig. 3(b)) by reducing striker velocity (Fig. 4).

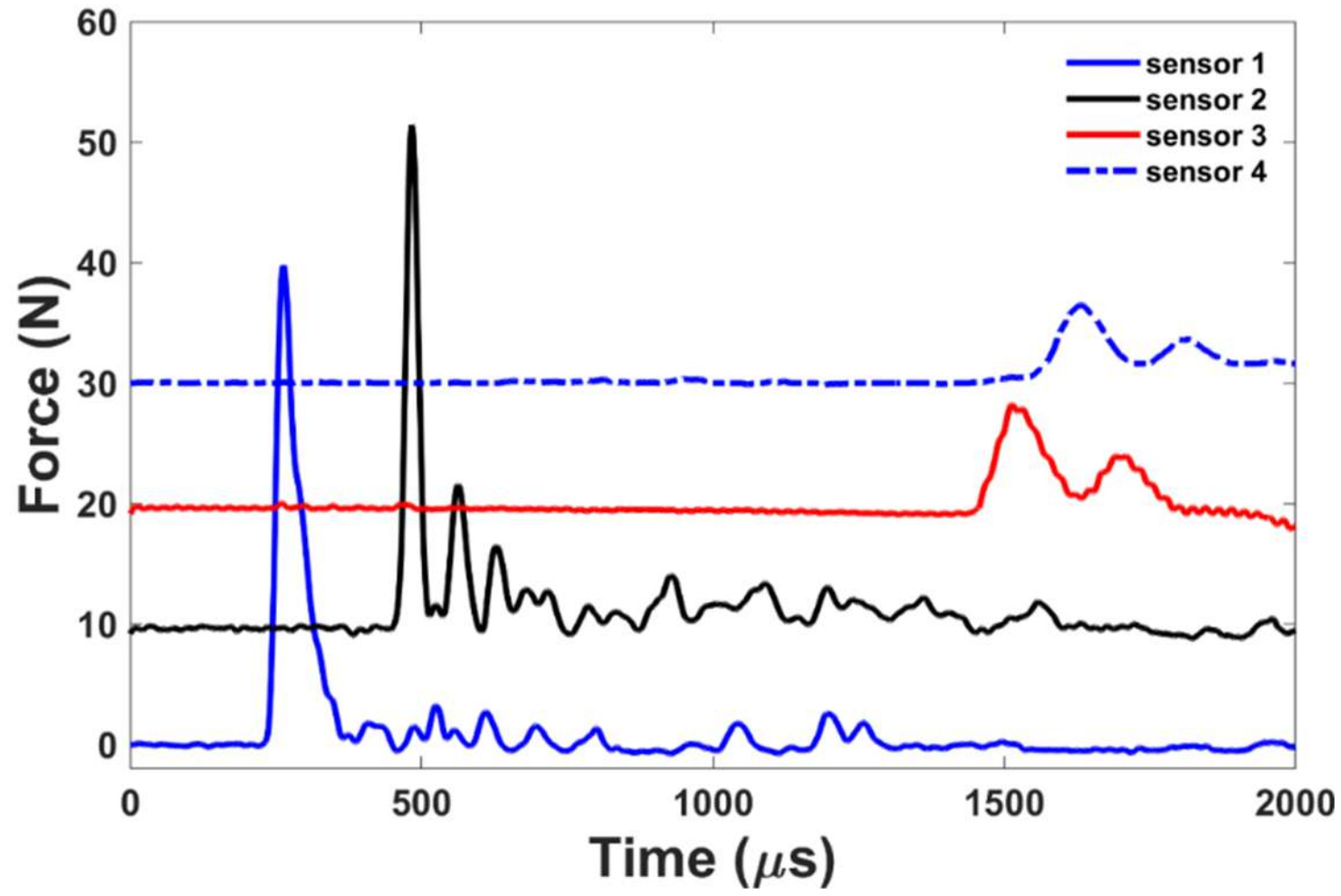

Figure 4. Solitary wave train generated by "hard" impact of striker with a lower velocity (0.4 m/s). The curves related to Sensors 1-4 are offset by 10 N for visual clarity.

Lower striker velocity in case of "hard" impact reduced the amplitude of the leading solitary wave (Fig. 4), making it comparable to the amplitude of the shock-like stress wave generated by "soft" impact presented in Fig. 3(b). But reducing the amplitude of the incoming pulse did not change the nature of the propagating wave – train of attenuating Nesterenko solitary waves in the case of "hard" impact.

It can be also noticed that shock-like stress waves detected by sensors 3 and 4 in case of "soft" impact with striker velocity 1 m/s (Fig. 3(b)) have amplitudes comparable to amplitude of leading solitary wave detected by sensor 2 in case in case of "hard" impact with velocity 0.4 m/s (Fig. 4).

**Discussion**

The standard way to generate single or multiple Nesterenko solitary waves in granular chains in most previous investigations was a "hard" striker impact on the chain [11], [21], [40-43] or explosion [44]. This paper presents the first experimental results comparing "hard" and "soft" impacts by an identical striker with the intention to explore how the ramp time of the incoming pulse can change the nature of the generated waves.

A few significant changes were observed in the nature and evolution of the incoming pulse between cases of direct impact of the striker ("hard" impact) and striker impact on a PTFE sphere inside a PTFE O-ring placed on top of the first cylinder ("soft" impact).

"Hard" impact generated a train of strongly nonlinear solitary waves like previous observations for an identical chain composed of stainless steel cylinders and spheres [14] and in chains composed of spheres [11].

"Soft" impact resulted in a dramatic increase in the ramp time of the incoming pulse to 250 microseconds in comparison with 30 microseconds in "hard" impact, as detected by Sensor 1 in Fig. 3(b) and Fig.3(a) correspondingly. Also, "soft" impact increased the total duration of incoming pulses to 700 microseconds in comparison with 230 microseconds in "hard" impact as detected by Sensor 1 (compare Fig. 3(b) and Fig. 3(a)). Incoming pulse in case of "soft" impact is close to bell shape with slight oscillations (Fig. 3(b)), while for "hard" impact incoming pulse is less symmetrical without oscillations (Fig. 3(a)).

"Soft" impact resulted in dramatically decreased amplitudes of incoming pulse (Fig. 3(b), Sensor 1), by almost a factor of five at the same velocity of the striker. It means that placing a small mass soft PTFE particle on the top of the chain (its mass was 84 times smaller than the striker mass) can be very beneficial in controlling the amplitude of the incoming pulse excited by the same striker impact.

The type of the propagated wave was dramatically changed from an attenuating train of solitary waves, Fig. 3(a) ("hard" striker impact), to an attenuating shock-like stress wave with small oscillations, Fig. 3(b) ("soft" striker impact). The latter has a typical shock-like profile with short duration, a stationary leading front and long tail, Fig. 3(b) ("soft" striker impact).

We emphasize that the same striker at the same velocity was used in both cases. This dramatic difference in wave response was observed in our experiments up to the 131st particle.

It was natural to expect that in the investigated weakly dissipative chain incoming pulse with a ramped profile excited by "soft" impact would steepen over short distances due to strong nonlinearity being subsequently transformed into the train of Nesterenko solitary waves or into an oscillatory shock-like stress wave. The physical origin of the train of solitary waves is related to the system's propensity to evolve into stationary waves, which are Nesterenko solitary waves for strongly nonlinear weakly dissipative discrete systems (they are quasi-stationary in a weakly dissipative case). The experimentally observed behavior only partially supports this expectation, demonstrating steepening of the leading front of the pulse approaching a duration like one-half the duration of the solitary wave and creating a shock-like stress wave with small oscillations (compare data from Sensors 1 and 2, Fig. 3(b)). But this shock-like stress wave did not develop into a train of solitary waves or into a highly oscillatory shock-like stress wave. Instead, it was transformed into an attenuating, slightly oscillating shock -like stress wave with a monotonic tail at investigated relatively long travel distances up to the 131st particle (Sensors 3 and 4, Fig.3(b).

Our experiments clearly demonstrated that solitary waves are generated by high gradients of stress in the front of the incoming pulse if the ramp time is short enough or comparable to the duration of solitary waves, as in Fig. 3(a).

Based on the evolution of shock-like stress wave (see data from Sensors 3 and 4, Fig. 3(b) it is reasonable to expect that this difference in wave response will be preserved even at larger distances. We attribute these changes in system behavior mainly to the ramping of the front part of the incoming pulse and not to the smaller amplitude or to the longer overall incoming pulse duration in “soft” impact.

This assumption can be verified by conducting “hard” impact experiments with incoming pulse amplitude similar to “soft” impact experiments. To generate amplitudes of the leading solitary wave, close to Fig. 3(b) for the ramped incoming pulse**,** a “hard” impact with the same striker having a smaller velocity, 0.4 m/s (striker dropped from 10 mm above the 1-D chain) was used. Results are shown in Fig. 4.

The prevailing influence of ramp time (not amplitude or dissipation) on the nature of propagating waves is evident by comparing Fig. 3 (b) and Fig. 4. Even at a smaller amplitude of the incoming pulse, the “hard” impact of the striker still results in a train of attenuating solitary waves. Thus, the main reason for generating a shock-like stress wave (non-solitary-type wave propagation) is the ramped nature of the incoming pulse front in “soft” impact (Fig. 3(b)) and not its smaller amplitude or increased influence of friction at a smaller amplitude of shock-like stress wave.

The ramp time, 250 microseconds (Sensor 1, Fig. 3(b)), which resulted in a shock-like stress wave, is about five times longer than the duration of the leading solitary wave in Fig. 2(a). It would be interesting to establish a criterion for the critical ramp time to secure a transition to a shock-like stress wave regime from solitary wave response. It is expected that this critical ramp time will be a few times larger than the solitary wave duration with an amplitude like the amplitude of the incoming pulse. This prediction is expected to be valid for other strongly nonlinear weakly dissipative ordered discrete systems.

Granular chains are a unique platform for theoretical and experimental investigation of strongly nonlinear discrete systems, opening an area of qualitatively new wave dynamics with a promising variety of applications. They support stationary strongly nonlinear Nesterenko solitary waves in non-dissipative systems, which are quasi-stationary in weakly dissipative systems. Strongly nonlinear shock-like stress waves are the only stationary waves in dissipative systems (oscillating or monotonous) generated in the piston problem, depending on the dissipation rate [45]. Observed regime change depending on the duration of incoming ramp time can be expected in ordered weakly dissipating 2-D and 3-D systems supporting propagation of quasi-stationary solitary waves. It is not expected for strongly nonlinear and strongly dissipative discrete ordered 1-D systems (e.g., systems with Nitrile O rings as nonlinear elements [47]) which do not support quasi-stationary Nesterenko solitary waves.

Compression waves generated by impact were investigated in disordered two and three-dimensional granular systems [48], [49]. They do not support solitary waves, unlike ordered one-dimensional or two/three-dimensional non-dissipative or weakly dissipative systems. Thus, it is not expected that a change of ramp time of the incoming pulse will result in a change of their responses from solitary wave type to shock-like stress wave observed in our article.

The effect of collision time between particles on different regimes of nonlinear wave propagation (from chain-like force propagation to dense force structure with a well-defined front) was observed in [48]. This change of response happened under the same "hard" impact condition when the ramp time of the incoming pulse (determined by the interaction between rigid intruder and particles) is like the collision time between particles. But in general collision time between particles and the ramp time of the incoming pulse are different parameters. Change of "hard" impact to "soft" impact and subsequently incoming ramp time in the experimental setup in [48] at the same velocity of the intruder can be accomplished by placing a soft layer at the top of the granular system with rigid particles. Change of ramp of time of incoming pulse caused by "soft" impact might change the regime of wave propagation along chain-like force in 2-D granular systems in the case of hard and intermediate particles, for cases when the value of nondimensional parameter *M'* is close or below 0.3 [48]. It can be expected that pulses propagating in force chains under "hard" impact are solitary waves.

We would like to mention that the condition that impacts is "soft" or "hard" depends on the relative duration of incoming pulse ramp time in comparison to duration of solitary waves in the chain. This duration depends on the elastic properties of particle material, their masses and the interaction laws between them. For example, the duration of solitary waves in a chain of PTFE beads [50] is a few hundred microseconds, depending on their amplitude, while in a chain of steel beads it is about 10 microseconds [11]. That is why the incoming pulse with ramp time as in "soft" impact test in this article, probably could be qualified as "hard" impact for the chain of PTFE particles.

The solitary pulse speed (e.g., for a pulse detected by Sensor 2, Fig. 3(a)) can be calculated using the distance between Sensors 1 and 2 and the time intervals between maxima of signals detected by these sensors. The following equation (Eq. (1)) to estimate Nesterenko solitary wave speed ($V_s$) was introduced in [14] for a similar chain of spheres and cylinders, which demonstrated a very good agreement with experiments

$$V_s \approx \frac{h+2R}{\sqrt{5}}\left(\frac{2ER^{1/2}}{3m^{3/2}(1-\nu^2)}\right)^{1/3} {F_m}^{1/6}\ , \qquad (1)$$

where $F_m$ is the maximum force in the solitary wave. This equation is like the equation for a solitary wave in a chain of spheres (Eq. 1.43 in [18]). Some modifications resulting in Eq.1 are due to a different constant in the Hertz law for contact between a sphere and the flat surface of a cylinder and considering that the distance between centers of neighboring particles $a$ in the sphere/cylinder chain is $a = (R + h/2)$, unlike $2R$ as in the sphere/sphere chain.

This equation (with elastic parameters $E$=193 GPa, n=0.29, typical for steels) was used to calculate the speed of the leading solitary wave detected by Sensor 2 in Fig. 3(a). Force amplitude in this pulse was equal to 186 N, resulting in a speed of 648 m/s from Eq. 1. Experimental speed (based on the time interval between peaks detected by Sensor 1 and Sensor 2 and their separating distance) was 624 m/s. A similar agreement between the predicted solitary wave speed using Eq. 1 and experimental data was observed in [14].

The equation (Eq. (2)) for shock-like stress wave speed ($V_{sh}$) in dissipative sonic vacuum assembled of spheres and cylinders, (which is equation 1.46(a) for a chain of spheres in [18]), modified in a similar way as the equation for non-dissipative solitary wave speed is

$$V_s \approx \frac{h+2R}{2}\left(\frac{2ER^{1/2}}{3m^{3/2}(1-\nu^2)}\right)^{1/3} {F_m}^{1/6} \; . \qquad (2)$$

It should be emphasized that scaling of dissipative shock-like speed with force amplitude is like scaling of non-dissipative Nesterenko solitary wave, despite their qualitative difference. Eq. 2 is valid at any level of dissipation, while the shock wave profile can be monotonous or oscillating depending on dissipation. In case of oscillating shock, the value of equilibrium force should be used instead of the maximum force at the front [18], [45], [46]. Using maximum force equal to 24 N in Eq. (2) (average between amplitudes of waves detected by Sensors 3 and 4) resulted in a speed value $V_{sh}$ = 515 m/s. The shock-like stress wave speed in experiments using signals detected by Sensors 3 and 4, Fig. 3(b), was calculated by dividing the distance between these sensors by the time intervals between maxima of monotonous wave profiles detected by these sensors. It resulted in a value of 532 m/s, which is in good agreement with the speed obtained using Eq. 2.

This study has a few limitations. For example, it did not present the quantitative value of the critical ramp time required for the change of system response from solitary train response to monotonous stress shock-like wave response, suggesting only an order of magnitude for the critical ramp time. Numerical modeling is desirable to find this critical value of ramp time using an experimentally validated weakly dissipative model for a similar granular chain [14]. The presented experimental setup was intentionally designed to be "friendly" for such numerical modeling.

**Conclusions**

Two types of dynamic loading were investigated: "hard" impact with direct striker impact on the system and "soft" impact where the striker was impacting top PTFE sphere placed inside PTFE O-ring on the top cylinder in chain. In the first case, the incoming pulse was split into a few Nesterenko solitary waves after traveling about twenty particles. In the case of "soft" impact, the incoming pulse generated by the striker did not split into a train of solitary waves despite initial steepening of the leading front due to strong nonlinearity. It propagated as a shock-like stress pulse for a long distance of 131 particles without splitting into a train of solitary waves. Thus, the generation of the train of solitary waves or shock-like stress waves in a weakly dissipative "sonic vacuum" under striker impact over short distances from the impacted end depends on the duration of the incoming pulse front. "Soft" impact dramatically reduced the maximum force of the propagating shock-like stress pulse in comparison with the amplitude of solitary waves generated by "hard" impact.

### Author contributions

Po-Hsun Chiu – carried out experimental research, prepared the original draft.

Vitali F. Nesterenko – Original idea, reviewed and revised the article.

### Declaration of Conflicting Interests

The authors declare that they have no known competing financial interests or personal relationships that could have influenced the work reported in this manuscript.

### Funding Sources

This research did not receive any specific grants from funding agencies in the public, commercial, or not-for-profit sectors.